\documentclass{amsart}
\usepackage[foot]{amsaddr} 

\usepackage{geometry}
\usepackage[numbers]{natbib}
\usepackage{doi}
\usepackage{graphicx}
\usepackage{threeparttable}
\usepackage{multirow}
\usepackage{multicol}
\usepackage{arydshln} 

\usepackage{amsmath}     
\usepackage{amsfonts}

\usepackage{caption}
\usepackage{subcaption}

\usepackage{cleveref}
\usepackage{bm}

\keywords{meta-analysis; risk of bias; robust Bayesian analysis; imprecise probability}

\begin{document}

\title[A robust Bayesian bias-adjusted random effects model]{A robust Bayesian bias-adjusted random effects model for consideration of uncertainty about bias terms in evidence synthesis}

\author{Ivette Raices Cruz$^1$}
\address{$^1$Centre for Environmental and Climate Science, Lund University, Lund, Sweden}
\email{ivette.raices\_cruz@cec.lu.se}

\author{Matthias C.M. Troffaes$^2$}
\address{$^2$Durham University, Department of Mathematical Sciences, UK}
\email{matthias.troffaes@durham.ac.uk}

\author{Johan Lindstr{\"o}m$^3$}
\address{$^3$Centre for Mathematical Sciences, Lund University, Lund, Sweden}
\email{johan.lindstrom@matstat.lu.se}

\author{Ullrika Sahlin$^1$}
\email{ullrika.sahlin@cec.lu.se}

\begin{abstract}
Meta-analysis is a statistical method used in evidence synthesis for combining, analyzing and summarizing studies that have the 
 same target endpoint and aims to derive a pooled quantitative estimate using fixed and random effects models or network models. 
Differences among included studies depend on variations in target populations (i.e. heterogeneity) and variations
in study quality due to study design and execution (i.e. bias). The risk of bias is usually
assessed qualitatively using critical appraisal, and quantitative bias analysis can be used to evaluate the influence of bias on the quantity of interest.
We propose a way to consider ignorance or ambiguity in how to quantify bias terms in a bias analysis by characterizing bias with imprecision (as bounds on probability) and use robust Bayesian analysis to estimate the overall effect. 
Robust Bayesian analysis is here seen as Bayesian updating performed over a set of coherent probability distributions, where the set emerges from a set of bias terms.
We show how the set of bias terms 
can be specified based on judgments on the relative magnitude of biases (i.e., low, unclear and high risk of bias)
in one or several domains of the Cochrane’s risk of bias table.
For illustration, we apply a robust Bayesian bias-adjusted random effects model to an already published meta-analysis 
on the effect of Rituximab for rheumatoid arthritis from the Cochrane Database of Systematic Reviews. 
\end{abstract}

\maketitle

\section{Introduction}\label{sec1}

Meta-analysis is a statistical method to combine, analyze and summarize the results of studies that have the 
same target endpoint and calculate 
a quantitative estimate of the overall effect (pooled treatment/intervention effect) 
using fixed and random effects models \citep{Higgins_2009, Cochrane_2019} or network models \citep{Dias_network_2013, Tengbin_2014}. 
There are several factors to consider in meta-analysis such as heterogeneity and bias. Heterogeneity, arising from between study variations in populations, interventions, exposures and outcome measures, can be considered when specifying the meta-analysis model. The potential of errors and biases (a.k.a. the risk of bias) due to differences in the design and conduct of the studies \citep{Higgins_2009, Turner_2009} is usually assessed qualitatively using critical appraisal \citep{Cochrane_2019, vanderBles_2019}.
The validity of the estimate of the overall effect in a meta-analysis is affected by the quality of data,  
and uncertainty associated with the model (and parameters within the model).
\Citet{vanderBles_2019} conclude that these two levels of uncertainty, which they refer to as indirect uncertainty (i.e. quality of the underlying knowledge e.g. expressed by a reflexive summary of our confidence in the models or the experts) and direct uncertainty (i.e. quantitative terms/expressions of uncertainty such as, a probability distribution or confidence interval) respectively, are usually communicated side by side.
These two levels of uncertainty are both relevant, but can be both confusing and difficult to combine when making decisions. Therefore approaches to turn indirect uncertainty into direct uncertainty are useful \citep{vanderBles_2019}. 

In practice, an analyst has the following alternatives to consider indirect uncertainty when characterizing direct uncertainty: 
1) to remove studies with a high risk of bias and conduct the analysis with the best available evidence (i.e. high quality studies),
2) to evaluate using sensitivity analysis the influence of including studies of lower quality in the meta-analysis, or
3) to include all (or a selection of) studies, but adjust for bias \citep{Dias_2013}. 
The last option can be carried out using quantitative bias analysis (or bias adjustment) 
 \citep{Lash_2009, Lash_2014, Stone_2020}. Quantitative bias analysis is a method that requires the meta-analysis model to be extended with bias adjustments (e.g. additive or proportional adjustments of study specific errors in the model) 
 \citep{Turner_2009, Spiegelhalter_2003,  Verde_2021}, and additional expert judgment  \citep{Turner_2009, vanderBles_2019, Rhodes_2020} on bias terms. 

Bias terms in quantitative bias analysis are different from parameters. Parameters in statistical models are fixed quantities that we are uncertain about, but want to estimate using statistical inference \citep{vanderBles_2019}. 
The bias terms are here treated as uncertain but fixed quantities that, in contrast to other parameters, we are not trying to learn. These bias terms are informed by expert judgment of detailed qualitative information on the design and execution of the studies.
In practice, experts may be ignorant about bias  
terms or struggle to specify them by single values. \Citet{Spiegelhalter_2003} explored different choices of bias terms using sensitivity analysis. \Citet{Turner_2009} elicited quantitative distributions for bias terms based on expert's judgments. 
\Citet{Verde_2021} considered bias terms as 
scale random variables, 
modeled by a probability distribution.
An alternative approach which avoids mixing uncertainties in bias terms and parameters, is to characterize  
ignorance or ambiguity about bias terms by a set of bias terms, 
and thus, it is not necessary to specify a unique value. 

The aim of this paper is to propose a way to consider 
uncertainty, arising from ambiguity or ignorance about bias terms, by modeling bias in a bias-adjusted random effects model with imprecision, and use robust Bayesian analysis to estimate the overall effect.  
Robust Bayesian analysis is here seen as Bayesian updating performed over a set of coherent probability distributions, resulting in a set of posterior distributions for the quantity of interest. In this case, the set of posteriors are the result of the set of bias terms. 
Hence, the suggested robust Bayesian bias analysis characterizes uncertainty in the overall effect by bounded probability \citep{Walley_1997,Coolen_2011,Sahlin_2021}, where the differences between bounds (i.e. the degree of imprecision) in the overall effect is a result of ambiguity or ignorance about the bias terms. In this way, it is possible to evaluate the impact of bias and the impact of uncertainty associated with the bias separately.

We use a bias-adjusted random effects model with an additive bias (
the study specific treatment effect is modeled as the sum of an overall effect, a study specific random effect and a study specific error) associated with each study specific effect (as was done in \citet{Turner_2009} and \citet{Spiegelhalter_2003}). 
We propose a way to specify the set of bias terms for all studies by considering 
information from the Cochrane risk of bias table (RoB table).  
For illustration, we apply a robust Bayesian bias-adjusted random effects model to a published meta-analysis 
about the effect of Rituximab for rheumatoid arthritis 
from the Cochrane Database of Systematic Reviews \citep{Lopez_Olivo_2015}.

In what follows, we first describe the robust Bayesian bias-adjusted random effects model (Section 2). Next, we propose 
how to specify a set of coherent bias terms (in the model referred to as study qualities) using qualitative judgments about study quality from the RoB table (Section 3).
We then present an application of robust Bayesian quantitative bias analysis to a systematic review about the effect of Rituximab (Section 4). 
We conclude with a discussion (Section 5). 

\section{Model specification}\label{sec2}

\subsection{A random effects model}

Consider a meta-analysis for binary outcome of K studies where $N_{ij}$ denotes the total number of patients in group $j$ ($j = 1$ Control and $j = 2$ Treatment) in study $i$ ($i = 1, \dots, K$) and $r_{ij}$ denotes the number of patients who have had a positive response respectively. 

The number of patients $r_{ij}$ can be modeled with a binomial distribution as follows:
\begin{equation}
	r_{ij} | p_{ij} \sim \mathrm{Binomial}(p_{ij}, N_{ij}).
\end{equation}
The probability of success $p_{ij}$ can be transformed using a link function (e.g. logit) as follows:
\begin{align}
\mathrm{logit}(p_{i1}) &= \ln \left(\frac{p_{i1}}{ 1 - p_{i1}}  \right) = \beta_{i}, \\
\mathrm{logit}(p_{i2}) &= \ln \left(\frac{p_{i2}}{ 1 - p_{i2}}\right) = \beta_{i} + \delta_{i},
\end{align}
where $\beta_{i}$ is the log-odds ratio (`log-OR') for the control group and $\delta_{i}$ is the specific treatment effect on the `log-OR' scale in study $i$.
The specific treatment effect being evaluated in the $i$th study, $\delta_i$, can be expressed as the sum of the overall effect, $\mu$, the study specific random effect, $\theta_i$ and study specific error, $\phi_i$: 
\begin{equation}
	\delta_{i} = \mu + \theta_i + \phi_i,
\end{equation}
on the `log-OR' scale.

In addition, uncertainty about $\theta_i$ and $\phi_i$ can be described by 
\begin{align}
	\theta_{i} | \sigma_{\theta}^2 & \sim \mathrm{Normal} (0, \sigma_{\theta}^2),  \\
	\phi_{i} & \sim \mathrm{Normal} (0, \sigma_{\phi_i}^2) \label{eq_phi}
\end{align}
where $\sigma_{\theta}^2$ represents between study 
variability and $\sigma_{\phi_i}^2$ is the variance of the bias (study specific error). Here, \cref{eq_phi} represents uncertainty in a single trial of a study. In repeated trials, which we are not modeling here, there may be correlation between repeated trials of the same study.
 We assume that the expectation of study specific random effect and study specific error, $\theta_{i}$ and $\phi_i$, is 0 because we do not 
consider that a random effect or bias would favour either the treatment or the control group. 
Then, uncertainty about $\delta_i$ can be represented by 
\begin{align}
	\delta_i | \mu, \sigma_{\theta}^2 & \sim \mathrm{Normal} (\mu, \sigma_{\theta}^2 + \sigma_{\phi_i}^2). \label{Eq: delta_RE_model}
\end{align}

\subsection{Bias adjustment model} 

To adjust for bias, \cref{Eq: delta_RE_model} is expressed as 
\begin{align}
	\delta_i | \mu, \sigma_{\theta}^2  & \sim \mathrm{Normal} \left(\mu , \frac{\sigma_{\theta}^2}{q_i}\right)
\end{align}
where $q_i = \frac{\sigma_\theta^2}{\sigma_\theta^2 + \sigma_{\phi_i}^2 }$ is a bias term that can be interpreted as the quality of 
study $i$ \citep{Turner_2009, Spiegelhalter_2003} (therefore, in what follows, bias terms are referred to as study qualities). 
Study quality represents how large a proportion of total variance is due to between study variability compared to bias/design uncertainty. 
The case $q_i = 1$ represents the situation where there is no bias whereas $q_i = 0.5$ implies equal between study variability and variance of the bias term for study $i$. 
For example, a high quality study is expected to have a small error in relation to the between study variability and should therefore have a relatively high study quality, e.g. $q=0.95$ (or other value close to 1). A low quality study should accordingly have a low quality,
e.g. $q = 0.1$ (or other value close, but strictly larger than zero).
The study qualities influence the relative contribution of each study to the likelihood of the model and hence the estimation of the overall effect. 

\subsection{Bayesian inference} 

To implement the model in a Bayesian framework, the following prior distributions are specified for the parameters 
\begin{subequations} \label{priors}
\begin{align}
\beta_i  &\sim \mathrm{Normal}(\mu_\beta,\sigma_\beta^2), \\
\mu  & \sim \mathrm{Normal} (\mu_\mu, \sigma_{\mu}^2), \\
\sigma^2_{\theta} &\sim \mathrm{InvGamma}(\alpha, \lambda)
\end{align}
\end{subequations}
where $\mu_\beta$, $\sigma^2_\beta$, $\mu_\mu$, $\sigma^2_\mu$, $\alpha$ and $\lambda$
are hyperparameters for the normal and inverse gamma prior distributions. The study qualities 
$q_i$ can be interpreted as hyperparmeters for the bias adjustment part, or as auxiliary parameters as they are not integrated out in the analysis. 
The Bayesian bias-adjusted random effects model is a probabilistic graphical model (\cref{fig: Bayesian  bias-adjusted model}) that represents  
probabilistic dependencies among data, parameters and hyperparameters.

\begin{figure}
\centerline{\includegraphics{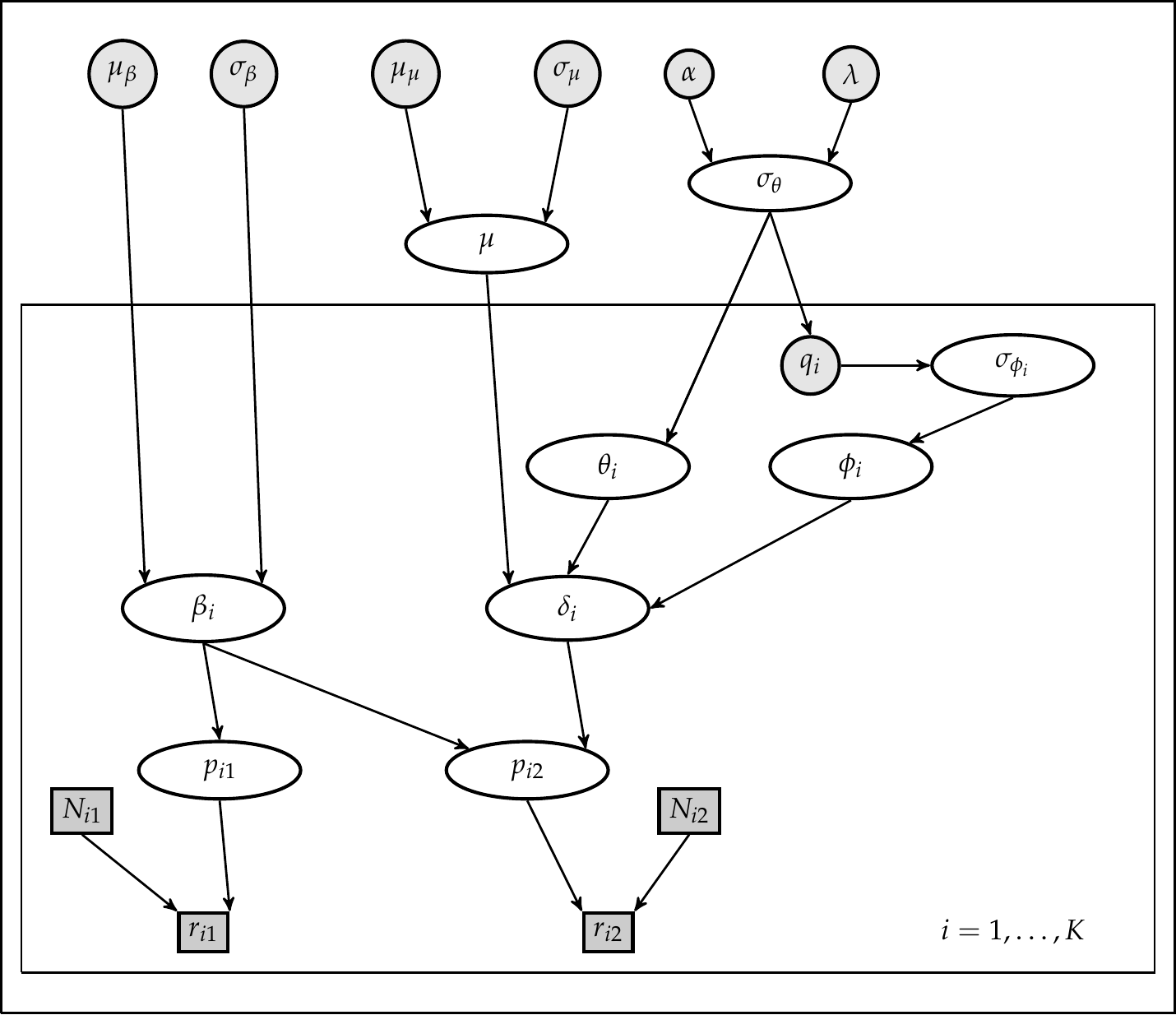}}
\caption{A probabilistic graphical representation of the Bayesian bias-adjusted random effects model. Unknown quantities (parameters) are represented by
white ellipses, for which priors are specified with fixed hyperparameters (gray circles). Observations (gray squares) are coming from K studies (the plate). The bias terms are fixed and therefore denoted by a gray circle.} \label{fig: Bayesian bias-adjusted model}
\end{figure}

\subsection{Robust Bayesian inference}

The Bayesian bias-adjusted random effects model is  
extended to 
a robust Bayesian framework by keeping the previously specified prior distributions and specifying a set of study qualities $\mathcal{M}$   resulting in:
\begin{align}
	\delta_i | \mu, \sigma_{\theta}^2  & \sim \mathrm{Normal} \left(\mu , \frac{\sigma_{\theta}^2}{q_i} \right) \mbox{ where } q_i \in \mathcal{M}.
\end{align}

The robust Bayesian bias-adjusted random effects model allows us to evaluate the impact of uncertainty about bias terms, $q_i$, on 
uncertainty of the quantity of interest (e.g. the overall effect of the intervention, $\mu$). 
This means that for different values $q_i \in \mathcal{M}$, there are different distributions for $\delta_i$ and thus, specifying a set of study qualities results in bounds on the probabilities that characterize uncertainty in the quantity of interest. 
The difference between lower and upper bounds for the probabilities expressing  uncertainty in the quantity of interest 
is therefore a result of ambiguity or ignorance about the bias terms in the quantitative bias analysis.

Uncertainty about the overall effect, $\mu$, is summarized by
bounds on the expected overall effect, bounds on the 5th percentile, and bounds on $P(\mu > t)$, the probability of the overall effect, $\mu$, exceeding a decision relevant threshold, $t\in\mathbb{R}$.  
For instance, the posterior expected overall effect $\mathrm{E}_{\mathbf{q}}(\mu)$ is calculated by 
a multiple integral of $\mu$ times the posterior distribution over the domains of the parameters $\beta_1, \beta_2, \beta_3, \beta_4,\delta_1, \delta_2, \delta_3, \delta_4, \mu,\sigma^2_\theta$ (see appendix for details). 
Then, lower and upper bounds on the expected overall effect are given by
\begin{align}
\underline{\mathrm{E}}_{\mathbf{q}}(\mu) &= \inf_{\mathbf{q} \in \mathcal{M}} \mathrm{E}_{\mathbf{q}}(\mu) \\
\overline{\mathrm{E}}_{\mathbf{q}}(\mu) &= \sup_{\mathbf{q} \in \mathcal{M}} \mathrm{E}_{\mathbf{q}}(\mu).
\end{align}

Bounds are here estimated using a discretization of the elements of the set of study qualities, in particular, 
we specify a finite set of study quality using a grid (details are given in Section 4). 
Then, for each study quality in this set, we estimate the expectation, percentile and exceedance probability of the overall effect using Markov chain Monte Carlo (MCMC) sampling which draws samples from the posterior distribution.
Finally, the lower and upper bounds are approximated by the minimum and the maximum values of each estimated quantity. 

\section{Quantification of study quality from the Cochrane's risk of bias table} \label{sec3}

We describe a way to incorporate qualitative judgments about risk of bias in a bias-adjusted meta-analysis. In particular, we show how to transform qualitative
judgments about risk of bias into quantitative expressions (i.e. a coherent set of study qualities) 
and thereby characterizing uncertainty in the specification of bias (study quality) by bounded probability. 

The Cochrane's risk of bias table (RoB table) is the recommended tool for assessing the risk of bias in each included study in Cochrane reviews \citep{Higgins_2011}. 
A RoB table takes into account the following domains: random sequence generation 
(selection bias), allocation concealment (selection bias), blinding of participants and personnel (performance bias), 
incomplete outcome data (attrition bias), selective outcome reporting (reporting bias) and other potential sources of 
bias \citep{Higgins_2011}. Each domain is assessed individually and classified in three categories: 
low, unclear and high risk of bias \citep{Higgins_2011}. Here, we describe a way to use this information that only requires a specification of the bounds for the lowest and highest value on the bias term $q_i$ across all studies $i$.

\subsection{Considerations for rating studies}
 
Let us consider a single domain of the RoB table for establishing a rating between the studies. Let $q_{(+)}$, $q_{(?)}$ and $q_{(-)}$ denote the study quality of studies with low, unclear and high  risk of bias, respectively. The approach is  explained using a hypothetical systematic review with 6 studies ($S_1, \dots, S_6$) which are classified as follows: $S_1$ and $S_2$ with a low risk of bias, $S_3$ and $S_4$ with an unclear risk of bias and $S_5$ and $S_6$ with a high risk of bias. 
The set of quantitative bias terms are constructed as follows: 
\begin{enumerate}
	\item Group the studies according to their risk of bias: low (+), unclear (?) and high (-). 

	\item All studies belonging to a group with a clear 
	risk of bias are assumed to have the same study quality.  
	For example: $q_1 = q_2 = q_{(+)}$ and $q_5 = q_6 = q_{(-)}$.

    \item The quality of studies belonging to the group with an unclear risk of bias are not necessarily equal and are somewhere in between 
		$q_{(-)} \leq q_{(?)} \leq q_{(+)}$. 
		For example: $q_{(-)} \leq q_3 \leq q_{(+)}$ and $q_{(-)} \leq q_4 \leq q_{(+)}$.
    
	\item Assign a range representing the lower and upper bounds for the bias terms in the low and high risk of bias categories. 
\end{enumerate}

In what follows, we let $q_{(+)}$ take values between 0.5 and 0.95, 
which corresponds to situations where between study variability contributes more to the total variance compared to the variance of study specific errors. 
We let $q_{(-)}$ take values between 0.1 and 0.5, 
which corresponds to situations where the variance of study specific errors contributes more than between study variability to the total variance. 
For the group of studies with unclear risk of bias, we use the most extreme bounds for the high and low risk of bias group, i.e. 
$0.1 = \underline{q_{(-)}} \leq q_{(?)} \leq \overline{q_{(+)}} = 0.95$.

\section{Application}

\subsection{Data} 

Data are taken from a systematic review about Rituximab for rheumatoid arthritis from the Cochrane Database of 
Systematic Reviews. 
For this example, we include 4 studies investigating the effect of Rituximab plus metrotexato (RT + MTX) vs placebo 
plus metrotexato (MTX) in patients with rheumatoid arthritis. The effect is assessed by the number of patients 
who have improved by at least 50\% on the American College of Rheumatology scale (ACR50) at week 
24 \citep{Lopez_Olivo_2015}. The outcomes of the studies and their assessment of risk of bias are shown in 
\cref{Tab: Info_studies} and \cref{Tab: RoB}, see  \citep[p.~15]{Lopez_Olivo_2015}, respectively.

  \begin{table}
    \centering
    \caption{Summary of studies \citep{Lopez_Olivo_2015}}
    \label{Tab: Info_studies} 
      \renewcommand{\arraystretch}{1.5}
    \begin{tabular}{|c | c | c | c | c |}
    \hline
    \multirow{4}{*}{Study Name} & \multicolumn{2}{|c|}{Control Group} & \multicolumn{2}{|c|}{Treatment Group}  \\ 
    &  \multicolumn{2}{|c|}{(MTX)} & \multicolumn{2}{|c|}{(RT + MTX)} \\ 
    \cline{2-5}
    & Response  & Total &  Response  & Total \\
    \hline
REFLEX (Study 1)  & 10 & 201 & 80 & 298  \\ \hdashline
WA16291 (Study 2) & 5 & 40 & 17 & 40  \\ \hdashline
DANCER (Study 3)  & 16 & 122 & 41 & 122 \\ \hdashline
SERENE (Study 4)  & 16 & 172 & 44 & 170 \\
    \hline
    Total  & 47 & 535 & 182 & 630 \\
    \hline
    \end{tabular}
  \end{table}

\begin{table}
		\centering
		  \begin{threeparttable}
		  \caption{Risk of bias table taken from the systematic review \citep{Lopez_Olivo_2015}}
		  \label{Tab: RoB}
		  \renewcommand{\arraystretch}{1.5}
		  \begin{tabular}{|c | c | c | c | c | c | c |}
		  \hline
		  Study Name & \bf{1} & \bf{2} & \bf{3} & \bf{4}  & \bf{5} & \bf{6} \\
		  \hline
		  REFLEX (Study 1)  & \smash{\raisebox{-1mm}{\includegraphics[scale =1]{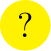}}} & \smash{\raisebox{-1mm}{\includegraphics[scale =1]{unclear_RoB_fig.pdf}}} & \smash{\raisebox{-1mm}{\includegraphics[scale =1]{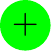}}} & \smash{\raisebox{-1mm}{\includegraphics[scale =1]{unclear_RoB_fig.pdf}}} & \smash{\raisebox{-1mm}{\includegraphics[scale =1]{low_RoB_fig.pdf}}} & \smash{\raisebox{-1mm}{\includegraphics[scale =1]{low_RoB_fig.pdf}}} \\ \hdashline
		  WA16291 (Study 2) & \smash{\raisebox{-1mm}{\includegraphics[scale =1]{unclear_RoB_fig.pdf}}} & \smash{\raisebox{-1mm}{\includegraphics[scale =1]{unclear_RoB_fig.pdf}}} & \smash{\raisebox{-1mm}{\includegraphics[scale =1]{low_RoB_fig.pdf}}} & \smash{\raisebox{-1mm}{\includegraphics[scale =1]{low_RoB_fig.pdf}}} & \smash{\raisebox{-1mm}{\includegraphics[scale =1]{low_RoB_fig.pdf}}} & \smash{\raisebox{-1mm}{\includegraphics[scale =1]{low_RoB_fig.pdf}}} \\ \hdashline
		  DANCER (Study 3)  & \smash{\raisebox{-1mm}{\includegraphics[scale =1]{unclear_RoB_fig.pdf}}} & \smash{\raisebox{-1mm}{\includegraphics[scale =1]{unclear_RoB_fig.pdf}}} & \smash{\raisebox{-1mm}{\includegraphics[scale =1]{unclear_RoB_fig.pdf}}} & \smash{\raisebox{-1mm}{\includegraphics[scale =1]{low_RoB_fig.pdf}}} & \smash{\raisebox{-1mm}{\includegraphics[scale =1]{low_RoB_fig.pdf}}} & \smash{\raisebox{-1mm}{\includegraphics[scale =1]{low_RoB_fig.pdf}}} \\ \hdashline
		  SERENE (Study 4)  & \smash{\raisebox{-1mm}{\includegraphics[scale =1]{unclear_RoB_fig.pdf}}} & \smash{\raisebox{-1mm}{\includegraphics[scale =1]{unclear_RoB_fig.pdf}}} & \smash{\raisebox{-1mm}{\includegraphics[scale =1]{unclear_RoB_fig.pdf}}} & \smash{\raisebox{-1mm}{\includegraphics[scale =1]{low_RoB_fig.pdf}}} & \smash{\raisebox{-1mm}{\includegraphics[scale =1]{low_RoB_fig.pdf}}} & \smash{\raisebox{-1mm}{\includegraphics[scale =1]{low_RoB_fig.pdf}}} \\ 
		  \hline
		  \end{tabular}
		  \begin{tablenotes}
			\item[{\bf 1}] random sequence generation (selection bias). 
		  	\item[{\bf 2}] allocation concealment (selection bias). 
		  	\item[{\bf 3}] blinding of participants and personnel (performance bias)
			\item[{\bf 4}] incomplete outcome data (attrition bias)
			\item[{\bf 5}] selective outcome reporting (reporting bias)
			\item[{\bf 6}] other potential sources of bias 
			\item[] \hspace{-4mm} \includegraphics[scale = 1]{low_RoB_fig.pdf} low risk of bias
			\vspace{1mm} 
			\item[] \hspace{-4mm} \includegraphics[scale = 1]{unclear_RoB_fig.pdf} unclear risk of bias
		  \end{tablenotes}
	  \end{threeparttable}
		  \end{table}

\subsection{Prior specification}

Noninformative prior distributions are used on the parameters $\beta_i, \mu$ and $\sigma^2_\theta$ which follow 
normal and inverse gamma distributions, \cref{priors}. The following  hyperparameter values 
are assigned: 
$\mu_\beta = 0$, $\sigma_\beta = 10$, $\mu_\mu = 0$, $\sigma_\mu = 10$, $\alpha = 0.01$ and 
$\lambda = 0.01$.
Let us note that the estimated overall effect of the Bayesian bias-unadjusted random effects model (\cref{Tab: RBA Outputs}) using these hyperparameter 
values gives estimates similar to the one, in the published Cochrane meta-analysis \citep{Lopez_Olivo_2015}.

\begin{table}
  \centering
    \begin{threeparttable}
	\caption{Bounds on expected value, exceedance probability and 5\% percentile of the overall effect, $\mu$, 
	 comparing Rituximab plus metrotexato (Treatment) against placebo plus metrotexato (Control) for robust bias adjusted meta analysis considering different groups of bias domains.
	$\mathbf{q_*}$ and $\mathbf{q^*}$ are the values for the bias terms where the bounds are obtained.
	The probability of the overall effect exceeding a threshold $t$ (where $t = 1$ for illustrative purposes)}
	\label{Tab: RBA Outputs} 
	\renewcommand{\arraystretch}{1.5}
	\begin{tabular}{|c |c | c | c | c | c |}
	\hline
	Bias & Quantity of & Lower & \multirow{2}{*}{$\mathbf{q_*}$} & Upper &  \multirow{2}{*}{$\mathbf{q^*}$} \\
	domain & interest & bound & & bound & \\
    \hline
    \multirow{3}{*}{\bf{1, 2} } &
    $\mathrm{E}(\mu)$  & 1.328  & (0.10, 0.10, 0.76, 0.95) & 1.646  & (0.95, 0.86, 0.10, 0.10)  \\
	& $P(\mu > 1)$ & 0.886 & (0.10, 0.10, 0.76, 0.76) & 0.983  & (0.95, 0.86, 0.10, 0.10) \\
	& $P_{5\%}$ & 0.826   & (0.10, 0.10, 0.19, 0.19) & 1.169 &  (0.95, 0.86, 0.10, 0.10) \\
	\hline
	\multirow{3}{*}{{\bf 3}} &  $\mathrm{E}(\mu)$  & 1.461 & (0.86, 0.86, 0.86, 0.78)  & 1.634  & (0.86, 0.86, 0.10, 0.10) \\
	& $P(\mu > 1)$ & 0.945  & (0.50, 0.50, 0.42, 0.50)  & 0.982 & (0.95, 0.95, 0.10, 0.10) \\
	& $P_{5\%}$ & 0.982 & (0.59, 0.59, 0.51, 0.59) & 1.159 &  (0.95, 0.95, 0.10, 0.10)  \\
	\hline
	\multirow{3}{*}{{\bf 4}} &  $\mathrm{E}(\mu)$  & 1.350  & (0.10, 0.91, 0.91, 0.91) & 1.476 & (0.59, 0.59, 0.59, 0.59) \\
	& $P(\mu > 1)$ & 0.902 & (0.10, 0.95, 0.95, 0.95) & 0.956 & (0.86, 0.86, 0.86, 0.86) \\
	& $P_{5\%}$ & 0.881 & (0.10, 0.59, 0.59, 0.59) & 1.025 & (0.82, 0.86, 0.86, 0.86)  \\
	\hline
	\multirow{3}{*}{{\bf 5, 6}} &  $\mathrm{E}(\mu)$  & 1.462 & (0.50, 0.50, 0.50, 0.50) & 1.478 & (0.90, 0.90, 0.90, 0.90) \\
	& $P(\mu > 1)$ & 0.945 & (0.50, 0.50, 0.50, 0.50) & 0.955 & (0.85, 0.85, 0.85, 0.85) \\
	& $P_{5\%}$ & 0.982 & (0.50, 0.50, 0.50, 0.50) & 1.020 & (0.85, 0.85, 0.85, 0.85) \\
	\hline
	\multirow{2}{*}{{\bf all}}  &  $\mathrm{E}(\mu)$  & 1.356 & (0.10, 0.95, 0.95, 0.95) & 1.638 & (0.87, 0.95, 0.10, 0.10) \\
	& $P(\mu > 1)$ & 0.905  & (0.10, 0.36, 0.36, 0.36)  & 0.982 & (0.87, 0.87, 0.10, 0.10)  \\
	{\bf 1, 2, 3 , 4, 5, 6} & $P_{5\%}$   & 0.847 & (0.10, 0.10, 0.10, 0.10) & 1.161 & (0.87, 0.87, 0.10, 0.10) \\
	\hline
	{\bf bias-}  &  $\mathrm{E}(\mu)$  & 1.471 &  & 1.471 &  \\
	{\bf unadjusted} & $P(\mu > 1)$ & 0.998 & -- & 0.998 & -- \\
	{\bf model} & $P_{5\%}$   & 1.029 & & 1.029 &   \\
	\hline
	\end{tabular}
	\begin{tablenotes}
		\item[{\bf 1}] random sequence generation (selection bias). 
        \item[{\bf 2}] allocation concealment (selection bias). 
        \item[{\bf 3}] blinding of participants and personnel (performance bias)
		\item[{\bf 4}] incomplete outcome data (attrition bias)
		\item[{\bf 5}] selective outcome reporting (reporting bias)
		\item[{\bf 6}] other potential sources of bias 
	\end{tablenotes}
	\end{threeparttable}
\end{table}

\subsection{Bias adjustment based on study quality}

Bias was accounted for by rating the studies according to each bias domain in the RoB table of the systematic review separately. For example, the effect of selection bias due to random sequence generation was evaluated separately from attrition bias due to incomplete outcome data. We illustrate each possible case according to each different domain listed in \cref{Tab: RoB}.  Specifically, we consider 4 cases corresponding to each of the 4 distinct columns in the table and an additional case corresponding to multiple domains.

\begin{itemize}
    \item {\bf Bias domains 1 and 2}    
    
The impact of selection biases due to random sequence generation and allocation concealment domains is similar 
because all the studies are rated with an unclear risk of bias. Following the  considerations in subsection 3.1: 

\begin{itemize}
\item Studies with unclear risk of bias can have different study qualities and satisfy 
\begin{equation}\label{case_3_q_domain_1}
0.1 \leq q_i \leq 0.95, \text{ } i = 1, \dots, 4
\end{equation}
\end{itemize}
which yields to the following set 
\begin{equation}
\mathcal{S}_1 = \mathcal{S}_2 =  \left\{\begin{array}{c}
										 \mathbf{q}:=(q_1, q_2, q_3, q_4): \\ 
										             0.1 \leq q_i \leq 0.95, \text{ } i = 1, \dots, 4 
				\end{array} \right \}. 	\label{Eq: set_S1_2}
\end{equation}

A regular grid of $10 \times 10 \times 10 \times 10$ of study quality $\mathbf{q}$  
is used for estimating the bounds, 
which results in 10 000 study quality values, $\mathbf{q}$.
More specifically, we considered $q_i$ equally spaced between 0.10 and 0.95.

\item {\bf Bias domain 3}  
    
Based on the impact of performance bias due to blinding of participants and personnel domain, the studies are grouped as follows: REFLEX (Study 1) and WA16291 (Study 2) in the category of low risk of bias and 
DANCER (Study 3) and SERENE (Study 4) in the category of unclear risk of bias. Then, we get: 
\begin{itemize}
\item Studies with low risk of bias have equal study qualities and studies with unclear risk of bias can have different study qualities and satisfy 
\begin{equation}\label{case_3_q_domain_3}
	q_3 \leq q_1, \text{ and }  q_4 \leq q_1 \text{ where }  q_1 = q_2. 
\end{equation}
\end{itemize}
Here, we make use of the previously established bounds for studies with low risk of bias and in the absence of studies with high risk of bias, $0.1$ is then used as the lower bound for studies with unclear risk of bias. The resulting set is

\begin{equation}
\mathcal{S}_3 =  \left\{\begin{array}{c}
										 \mathbf{q}:=(q_1, q_2, q_3, q_4): \\ 
										q_2 = q_1,  \\
	                                    0.1 \leq q_3 \leq q_1, \\
	                                    0.1 \leq q_4 \leq q_1, \\
	                                    0.5 \leq q_1 \leq 0.95 
				\end{array} \right \}	\label{Eq: set_S3}
\end{equation}
For estimating the bounds, we use a discretization of the convex combination of the extreme points of this set which can be seen as the weighted sum of the extreme points of the set. 
In details, these are the extreme points of $\mathcal{S}_3$:
\begin{align}
\mathbf{v_1} &= (0.50, 0.50, 0.10, 0.10), \nonumber \\
\mathbf{v_2} &= (0.95, 0.95, 0.10, 0.10), \nonumber \\
\mathbf{v_3} &= (0.50, 0.50, 0.10, 0.50), \nonumber \\
\mathbf{v_4} &=  (0.95, 0.95, 0.10, 0.95), \nonumber \\
\mathbf{v_5} & = (0.50, 0.50, 0.50, 0.10),  \nonumber \\
\mathbf{v_6} &= (0.95, 0.95, 0.95, 0.10), \nonumber \\
\mathbf{v_7} &= (0.50, 0.50, 0.50, 0.50), \nonumber \\
\mathbf{v_8} &= (0.95, 0.95, 0.95, 0.95), \nonumber 
\end{align}
and study qualities are specified as 
\begin{equation}
  \mathbf{q} = \sum_{i = 1}^8 \alpha_i \cdot \mathbf{v_i} \text{ where } \sum_{i = 1}^8 \alpha_i = 1 \text{ and } \alpha_i \geq 0. \label{generated_study_quality_vals}
\end{equation}
In \cref{generated_study_quality_vals}, $\alpha_i$ represents the weights. 
We used a grid spacing of 0.1 for each $\alpha_i$ ($i = 1,\dots,7)$ for estimating the bounds. Specifically, the following values are used:
\begin{align}
    \alpha_1 & \in \{0, ~ 0.1, \dots, ~ 1\}, \nonumber \\
    \alpha_{k} & \in \left\{0, \dots, 1 -  \sum_{i = 1}^{k-1} \alpha_i \right\}, \qquad k=2,\ldots,7 \nonumber \\
    \alpha_8 & = 1 - \sum_{i = 1}^7 \alpha_i.  \nonumber
\end{align}
Duplicate values of $\mathbf{q}$ are removed, which results in 736 study quality values, $\mathbf{q}$.

\item {\bf Bias domain 4}   

Based on attrition bias due to incomplete outcome data, the studies are grouped as follows: 
WA16291 (Study 2), DANCER (Study 3) and SERENE (Study 4) in the category of low risk of bias and  REFLEX (Study 1) 
in the category of unclear risk of bias. Then, we get: 
\begin{itemize}
\item Studies with unclear risk of bias can have different study qualities and satisfy 
\begin{equation}\label{case_3_q_domain_4}
	q_1 \leq q_2,  \text{ where }  q_2 = q_3 = q_4, 
\end{equation}
\end{itemize}
similarly, we get
\begin{equation}
\mathcal{S}_4 =  \left\{\begin{array}{c}
										 \mathbf{q}:=(q_1, q_2, q_3, q_4): \\ 
										q_2 = q_3 = q_4,  \\
	                                    0.1 \leq q_1 \leq q_2, \\
	                                    0.5 \leq q_2 \leq 0.95 
				\end{array} \right \}. 	\label{Eq: set_S4}
\end{equation}

Following the same procedure as in domain 3, the extreme points of $\mathcal{S}_4$ are: 
\begin{align}
\mathbf{v_1} & = (0.10,0.95,0.95,0.95), \nonumber \\
\mathbf{v_2} &= (0.10,0.50,0.50,0.50), \nonumber \\ 
\mathbf{v_3} &= (0.50,0.50,0.50,0.50), \nonumber \\ 
\mathbf{v_4} &= (0.95,0.95,0.95,0.95). \nonumber
\end{align}
Using a grid spacing of 0.1 of $\alpha_i$ (weights) we used the following values 
\begin{align}
    \alpha_1 & \in \{0, ~ 0.1,\dots, ~ 1\}, \nonumber \\
    \alpha_{k} & \in \left\{0, \dots, 1 -  \sum_{i = 1}^{k-1} \alpha_i \right\}, \qquad k=2,3 \nonumber \\
     \alpha_4 & = 1 - \sum_{i = 1}^{3} \alpha_i. \nonumber
\end{align}
The weighted sum of the extreme points of this set yields 286 study quality values, $\mathbf{q}$.

\item {\bf Bias domains 5 and 6} 

The impact of bias due to selective outcome reporting and other potential sources of bias domains are similar because all the studies are rated with a low risk of bias. Thus: 
\begin{itemize}
\item All the studies in the group of low risk of bias have the same study quality
\begin{equation}\label{case_3_q_domain_5_6}
q_1 = q_2 = q_3 = q_4, 
\end{equation}
\end{itemize}
and therefore
\begin{equation}
\mathcal{S}_5 = \mathcal{S}_6 =   \left\{\begin{array}{c}
										 \mathbf{q}:=(q_1, q_2, q_3, q_4): \\ 
										q_1 = q_2 = q_3 = q_4,  \\
	                                    0.5 \leq q_1 \leq 0.95 
				\end{array} \right \}. 	\label{Eq: set_S5_6}
\end{equation}

In this case, 10 equally spaced values between 0.5 and 0.95 of $q_1$ are considered, which results in 10 study quality values,  $\mathbf{q}$.

\item {\bf Multiple bias domains}

So far, we have focused on single domains of the RoB table. However, the proposed methodology could be extended or 
adapted, by making use of more than one domain of the risk of bias table. This requires a clear and transparent 
guidance on how to rate studies considering quality of data from several domains at once, as well as on how to prioritize domains. 

An example on how multiple bias domains could be combined for rating the studies is presented. 
The following considerations are taken into account:

\begin{itemize}
    \item WA16291 (Study 2) should be better than the rest of the studies, (it has 4 domains with a low risk of bias and 2 with an unclear risk of bias).
    
    \item DANCER (Study 3) and SERENE (Study 4) have the same type of biases
    
    \item It is not possible to relate REFLEX (Study 1) to DANCER (Study 3) or SERENE (Study 4) (we do not know if domain 3 and 4 are equally bad or if one is better than the other) 
\end{itemize}
which results in:
\begin{equation} \label{case_all_domains}
	q_1 \leq q_2,  \text{ } q_3 \leq q_2  \text{ where }  q_3 = q_4. 
\end{equation}

In this case, it is not possible to arrive to a final category of risk of bias without introducing a scoring rule (since all studies have domains with low and unclear risk of bias). 
Therefore, once again, we specify the most extreme bounds to account for uncertainty, 
in the specification of bias (study quality), resulting in:
\begin{equation}
\mathcal{S}_{\text{all}} =  \left\{\begin{array}{c}
										 \mathbf{q}:=(q_1, q_2, q_3, q_4): \\ 
										q_3 = q_4,  \\
	                                    0.1 \leq q_1 \leq q_2, \\
	                                    0.1 \leq q_3 \leq q_2, \\
	                                     0.1 \leq q_2 \leq 0.95 
				\end{array} \right \}. 	\label{Eq: set_all}
\end{equation}
The same procedure as in domains 3 and 4 is followed. The extreme points of $\mathcal{S}_{\text{all}}$ are: 
\begin{align}
\mathbf{v_1} &= (0.10, 0.10, 0.10, 0.10), \nonumber \\
\mathbf{v_2} &= (0.10, 0.95, 0.10, 0.10), \nonumber \\
\mathbf{v_3} &= (0.10, 0.95, 0.95, 0.95), \nonumber \\
\mathbf{v_4} &= (0.95, 0.95, 0.10, 0.10), \nonumber \\
\mathbf{v_5} &= (0.95, 0.95, 0.95, 0.95). \nonumber 
\end{align}
We use a grid spacing of 0.1 of $\alpha_i$, specifically, 
\begin{align}
    \alpha_1 & \in \{0, ~ 0.1,\dots,~ 1\}, \nonumber \\
    \alpha_{k} & \in \left\{0, \dots, 1 -  \sum_{i = 1}^{k-1} \alpha_i \right\}, \qquad k=2,\ldots,4 \nonumber \\
    \alpha_5 & = 1 - \sum_{i = 1}^{4} \alpha_i, \nonumber 
\end{align}
which results in 839 study quality values, $\mathbf{q}$.
\end{itemize}

\subsection{Estimation of treatment effect adjusted for bias}

\begin{figure}
      \centering
      \includegraphics[width=0.95\textwidth]{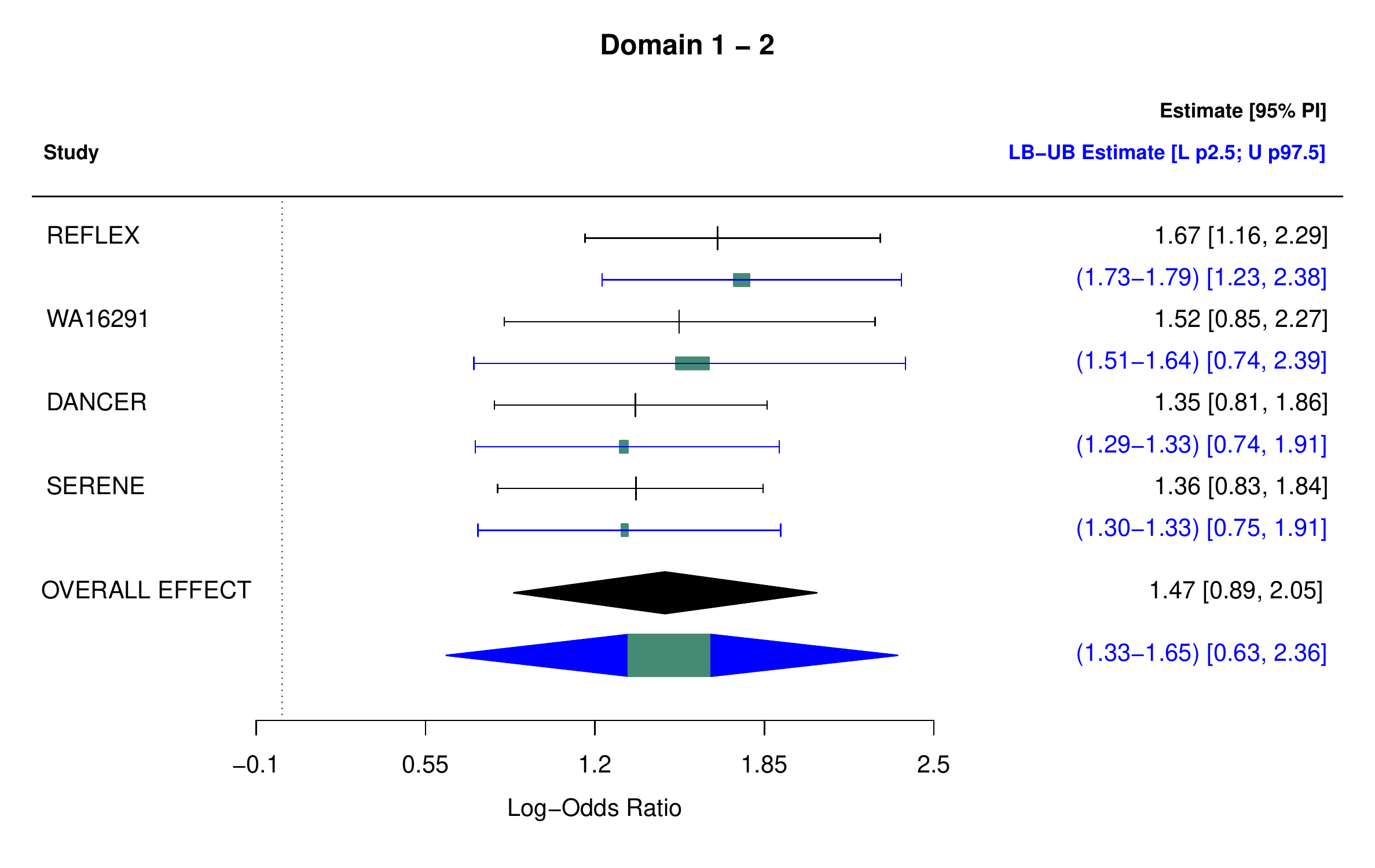}\label{fig:Forestplot_D1}
      \caption{Forestplot of a meta-analysis of the effectiveness of Rituximab plus metrotexato modified to show bounds on quantities of interest. Unadjusted and robust Bayesian bias-adjusted random effects log-odds ratios (with 95\% PI) are displayed: (black) unadjusted model; (blue) robust bias-adjusted random effects model. For the robust bias-adjusted random effects model, bounds on the expected overall effect, the lower 2.5th percentile and the upper 97.5th percentile are shown. 
}\label{fig:Forestplot_D1_only}
\end{figure}

Study specific treatment effects and overall effect are estimated using a robust Bayesian bias-adjusted random effects model.
The model is implemented using MCMC sampling in JAGS through the \emph {rjags} and \emph {runjags} packages, which are R interfaces to JAGS (see supplementary material for the code). 
Forestplots (a common graphical representation in meta-analysis) are used to present the estimated overall effect and specific study effect using the \emph {metafor} package for R. 
To express the results from robust Bayesian analysis, the forestplots from \emph {metafor} were expanded with bounds on the expected overall effect, $\mathrm{E}(\mu)$, the lower 2.5th percentile and the upper 97.5th percentile (\cref{fig:Forestplot_D1_only}). The results of the random effects model with and without robust Bayesian bias-adjustment  are 
displayed for each risk of bias domain in the appendix. 

\begin{figure}
      \centering
      \includegraphics[width=0.95\textwidth]{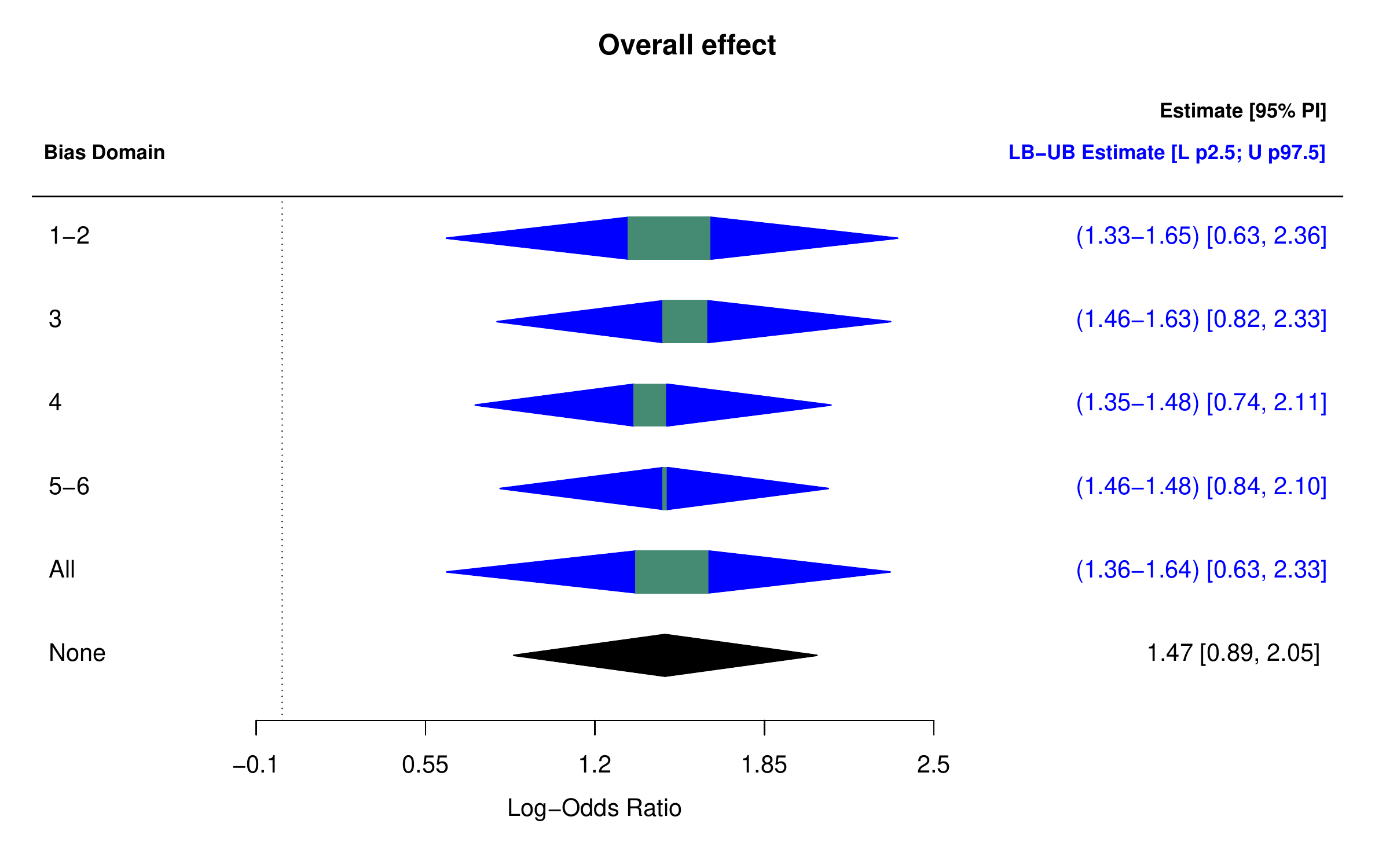}\label{fig:Overall_effect_plots}
   \caption{Uncertainty in the overall effect per bias domain.
   For the robust bias-adjusted random effects model, lower and upper bounds on the expected overall effect, a lower bound on the 2.5th percentile and an upper bound on the 97.5th percentile are shown. }\label{fig: Overall_effect_plots}
\end{figure}

Adjusting for bias associated with bias domains 1 and 2 changed the estimated overall effect from 1.47 to between 1.33 and 1.65; and resulted in a lower bound of the two sided 95\% probability interval (i.e. lower 2.5th percentile) to change from 0.89 to 0.63 (\cref{fig:Forestplot_D1_only}).
The difference between the lower and upper bound (i.e. the degree of imprecision) of the expected overall effect varied when adjusting for different risk of bias domains (\cref{fig: Overall_effect_plots}). 
As expected, there is more imprecision in the estimated overall effect, when all studies have an unclear risk of bias (Domain 1-2). For bias domains 5 and 6, all studies have a low risk of bias 
and were, according to the considerations set up for constructing the study quality sets, given equal bias terms. 
As a consequence, the between study variability has a very high influence on the total variance compared to the variance of bias, 
and therefore the overall effect is very similar to the unadjusted model. 
For domains 3 and 4, all studies, except WA16291 (Study 2) swapped their risk of bias category, between low and unclear, respectively. 
This explains why the overall effect is adjusted upwards (Domain 3) and downwards (Domain 4) compared to the unadjusted case. 

For a given domain, the values where the bounds are obtained, $\mathbf{q_*}$ and $\mathbf{q^*}$, can be different depending on the quantity of interest (\cref{Tab: RBA Outputs}).
For domains 1 and 2, the lower bound for the expected overall effect is obtained at $\mathbf{q_*}=(0.10, 0.10, 0.76, 0.95)$ whereas the lower bound for the probability of the overall effect exceeding 1 is obtained at 
$\mathbf{q_*} = (0.10, 0.10, 0.76, 0.76)$. The values of $\mathbf{q_*}$ and $\mathbf{q^*}$ are not necessarily on the extreme points of the set. Hence, it may be difficult to know in advance the values of $\mathbf{q}$ where the bounds are obtained.

Adjusting for bias may reveal important aspects to consider when framing a conclusion in evidence synthesis. Here, the evidence in favor of Rituximab plus metrotexato (treatment group) from the published meta-analysis \citep{Lopez_Olivo_2015} remains strong after adjusting for bias. We can see that the lower 2.5th percentile of the 
overall effect do not cross the reference value zero, for all risk of bias domains (\cref{fig: Overall_effect_plots}). Thus, the conclusion is robust to uncertainty (including ambiguity of the specification of bias terms). 

\section{Discussion and conclusion}\label{sec5}

Quantitative bias analysis \citep{Lash_2009, Lash_2014} is a statistical approach to combine direct and indirect uncertainty in evidence synthesis. Indirect uncertainty is often given as qualitative judgments on risk of biases \citep{vanderBles_2019}. Hence, bias analysis requires qualitative judgments of risk of bias to be transformed into quantitative expressions of uncertainty associated with estimates from studies (bias terms). In practical applications, it can be challenging to come up with precise bias terms. In this paper, we therefore propose robust Bayesian bias analysis as a way to consider
ambiguity or ignorance about how to quantify bias terms in meta-analysis, 
that distinguishes the impact of uncertainty about the bias terms from other uncertainties in the model. 
The proposed approach is a novel way to, in a structured way, use qualitative information concerning risk of biases in quantitative bias analysis, and bridge the gap between qualitative and quantitative expressions of uncertainty.
This is done by characterizing uncertainty about the bias terms by a set of bias terms, whereas parameter uncertainty is expressed by subjective probability. These two types of uncertainty are sometimes referred as second and first order uncertainty respectively \citep{Hansson_2008}. 
First order uncertainty is seen as the ``classical" Bayesian idea of unknown components that we marginalize over to obtain a posterior distribution conditional on only data, while the second order uncertainty is a set of possible fixed parameters for which we want to find a best or worst case of the posterior.
Characterizing second order uncertainty by a set of bias terms results in bounds on the probability representing first order uncertainty. 
Bounded probability is a suitable expression of uncertainty to represent expert's knowledge in situations of ignorance or ambiguity \citep{Walley_1997, Coolen_2011}. In the suggested method, second order uncertainty quantified within a model is seen as a difference between lower and upper bounds in quantities summarizing uncertainty about the overall effect. 

The proposed framework is not limited to the RoB table. It is still valid and useful, if a different risk of bias table is used. However, modifications may be necessary particularly if there are more than three risk of bias categories. In general, we need to 
i) decide the different categories of risk, 
ii) assign a category of risk to each study and
iii) specify the bounds per risk category. All steps are done on a case by case basis.
In this paper, we used the RoB table because it is the recommended and therefore, most used tool to assess risk of bias in randomized clinical trials.
The proposed framework makes use of the risk of bias table, so either the risk of bias table or expert's bias judgments should be available. 

The suggested approach estimates quantities of interest using robust Bayesian analysis, which gives more conservative estimates of a quantity of interest compared to standard Bayesian analysis.
It is, on the other hand, a more complex model to set up and usually requires the use of numerical algorithms to approximate the bounds. Bounds over the set of study qualities are in this paper approximated using a grid search approach through a discretization of the set, where Bayesian inference is done for each choice of study qualities.
Additionally, an increase in the number of studies may affect the computational burden of the discretization of the elements of the set for estimating the bounds. This will be particularly so when all studies have an unclear risk of bias, in which case a larger space must be explored.
An alternative approach is to search for bounds and do Bayesian updating in an iterative way using iterative importance sampling \citep{Troffaes_2017, Troffaes_2018}.

The choice of cut-off values of study quality $q$, is subjective. That is why, our selection, ($0.1, ~0.5,~ 0.95$), is quite conservative and includes most of the cases, as well as avoids possible numerical problems when values of $q$ are too close to zero. The mid cut-off value (0.5) is chosen because it is the median between 0 and 1 and links both categories of risk: high and low risk of bias. In this paper, we grouped the studies based on three different categories of risk of bias and then we specified possible values for each group. 
In practice, cut-off values of study quality $q$ can be informed from external information.
For instance, future research, through meta-epidemiological studies (analysis of multiple meta-analysis), similar case studies in combination with expert knowledge elicitation  \citep{Turner_2009, Rhodes_2020, Welton_2009}, may provide empirical results to motivate less conservative 
choices of cut-off values of study quality $q$. 

An added value of robust Bayesian bias analysis, compared to a standard Bayesian bias analysis
(i.e. Bayesian bias analysis using a single value or a precise probability distribution of study quality) is that it can show that the conclusion is not affected by risk of bias as well as by ignorance or ambiguity regarding how to quantify the bias terms. 
Robust Bayesian bias analysis can be performed as a first and coarse step motivating a refined bias analysis. When adjusting for bias does (not) have a high impact on the conclusion of the meta-analysis and therefore, a more detailed analysis may (not) be needed. A more detailed analysis could for instance gather more information regarding risk of bias or more carefully elicit bounds on study qualities. It can also be used to assess the influence of risk of bias on the conclusion of the assessment and then, if necessary, perform a Bayesian bias analysis relying on expert knowledge elicitation to specify a precise value or distribution of study quality.
In the application, the bounds from adjustment made on risk of biases categorized as unclear for all studies (such as Domain 1-2) contain the bounds from any other judgment of risk of biases (\cref{fig: Overall_effect_plots}).
Hence, robust Bayesian bias adjustment with unclear risk of biases is the most extreme case scenario towards a more refined approach. Robust adjustment with unclear risk of bias can be done without explicit judgments on risk of bias to explore if a quantitative bias adjustment may have a high impact on the result from the meta-analysis. 
Consider, as an example, a decision maker that will approve a treatment if certainty that the overall effect is larger than $t$ (for illustrative purposes the threshold $t$ is set to be $1$ in the application) is at least 95\%. The unadjusted estimate in the application is 0.998 (\cref{Tab: RBA Outputs}), and the decision maker wants to know if adjusting for bias may have an impact on the conclusion. Robust Bayesian bias adjustment with unclear risk of bias for all studies reveals that this probability can be as low as 0.886. Since this is below 95\%, it can be worthwhile to do a refined bias adjustment. 

To summarize, the proposed approach provides a structured framework to consider ambiguity or ignorance in quantitative bias analysis. 


\section*{Supplementary material}

The JAGS and R codes to run the analysis are available through the following link:
\url{https://github.com/Iraices/Robust\_bias\_adjustment}


\subsection*{Author contributions}

Conceptualization and Methodology IRC, US, MT and JL. Coding, IRC and
US. Formal analysis, Visualization and Writing - Original Draft, IRC; Writing
- Review and Editing, IRC, US, MT and JL.

\section*{Funding}

This work was supported by the Swedish research council FORMAS through the project 
``Scaling up uncertain environmental evidence'' (219-2013-1271) and the strategic research areas BECC 
(Biodiversity and  Ecosystem Services in a Changing Climate) and MERGE (Modelling the Regional and Global 
Climate/Earth system).

\subsection*{Conflict of interest}

The authors declare no potential conflict of interests.

\bibliographystyle{plainnat}
\bibliography{references_3}

\begin{thebibliography}{22}
\providecommand{\natexlab}[1]{#1}
\providecommand{\url}[1]{\texttt{#1}}
\expandafter\ifx\csname urlstyle\endcsname\relax
  \providecommand{\doi}[1]{doi: #1}\else
  \providecommand{\doi}{doi: \begingroup \urlstyle{rm}\Url}\fi

\bibitem[Coolen et~al.(2011)Coolen, Troffaes, and Augustin]{Coolen_2011}
Frank P.~A. Coolen, Matthias C.~M. Troffaes, and Thomas Augustin.
\newblock Imprecise probability.
\newblock In Miodrag Lovric, editor, \emph{International Encyclopedia of
  Statistical Science}, pages 645--648. Springer Berlin Heidelberg, 2011.
\newblock ISBN 978-3-642-04898-2.
\newblock \doi{10.1007/978-3-642-04898-2_296}.

\bibitem[Dias et~al.(2013{\natexlab{a}})Dias, Sutton, Ades, and
  Welton]{Dias_network_2013}
Sofia Dias, Alex~J. Sutton, A.~E. Ades, and Nicky~J. Welton.
\newblock Evidence {S}ynthesis for {D}ecision {M}aking 2: {A} {G}eneralized
  {L}inear {M}odeling {F}ramework for {P}airwise and {N}etwork {M}eta-analysis
  of {R}andomized {C}ontrolled {T}rials.
\newblock \emph{Medical {D}ecision {M}aking}, 33\penalty0 (5):\penalty0
  607--617, 2013{\natexlab{a}}.
\newblock \doi{10.1177/0272989X12458724}.

\bibitem[Dias et~al.(2013{\natexlab{b}})Dias, Sutton, Welton, and
  Ades]{Dias_2013}
Sofia Dias, Alex~J. Sutton, Nicky~J. Welton, and A.E. Ades.
\newblock Evidence synthesis for decision making 3: {H}eterogeneity-subgroups,
  meta-regression, bias, and bias-adjustment.
\newblock \emph{Medical Decision making}, 33\penalty0 (5):\penalty0 618--640,
  2013{\natexlab{b}}.
\newblock \doi{10.1177/0272989X13485157}.

\bibitem[Hansson(2008)]{Hansson_2008}
Sven~Ove Hansson.
\newblock Do we need second-order probabilities?
\newblock \emph{Dialectica}, 62\penalty0 (4):\penalty0 525--533, 2008.
\newblock ISSN 00122017, 17468361.

\bibitem[Higgins et~al.(2009)Higgins, Thompson, and
  Spiegelhalter]{Higgins_2009}
Julian P.~T. Higgins, Simon~G. Thompson, and David~J. Spiegelhalter.
\newblock A re-evaluation of random-effects meta-analysis.
\newblock \emph{Journal of the Royal Statistical Society}, 172\penalty0
  (1):\penalty0 137--159, 2009.

\bibitem[Higgins et~al.(2011)Higgins, Altman, Gotzsche, Jueni, Moher, Oxman,
  Savovic, Schulz, Weeks, Sterne, Grp, and Grp]{Higgins_2011}
Julian P.~T. Higgins, Douglas~G. Altman, Peter~C. Gotzsche, Peter Jueni, David
  Moher, Andrew~D. Oxman, Jelena Savovic, Kenneth~F. Schulz, Laura Weeks,
  Jonathan A.~C. Sterne, Cochrane Bias~Methods Grp, and Cochrane Stat~Methods
  Grp.
\newblock The {C}ochrane {C}ollaboration's tool for assessing risk of bias in
  randomised trials.
\newblock \emph{BMJ-BRITISH MEDICAL JOURNAL}, 343, October 2011.
\newblock \doi{10.1136/bmj.d5928}.

\bibitem[Higgins et~al.(2019)Higgins, Thomas, Chandler, Cumpston, Li, Page, and
  Welch]{Cochrane_2019}
Julian P.~T. Higgins, James Thomas, Jacqueline Chandler, Miranda Cumpston,
  Tianjing Li, Matthew~J. Page, and Vivian~A. Welch, editors.
\newblock \emph{Cochrane Handbook for Systematic Reviews of Interventions
  version 6.0 (updated July 2019)}.
\newblock Wiley-Blackwell, 2019.
\newblock URL \url{www.training.cochrane.org/handbook}.

\bibitem[Lash et~al.(2009)Lash, Fox, and Fink]{Lash_2009}
Timothy~L. Lash, Matthew~P. Fox, and Aliza~K. Fink.
\newblock \emph{Applying Bias Analysis to Epidemiologic Data}.
\newblock Springer, 2009.

\bibitem[Lash et~al.(2014)Lash, Fox, MacLehose, Maldonado, McCandless, and
  Greenland6]{Lash_2014}
Timothy~L. Lash, Matthew~P Fox, Richard~F. MacLehose, George Maldonado,
  Lawrence~C. McCandless, and Sander Greenland6.
\newblock Good practices for quantitative bias analysis.
\newblock \emph{International Journal of Epidemiology}, 43\penalty0
  (6):\penalty0 1969–1985, 2014.
\newblock \doi{10.1093/ije/dyu149}.

\bibitem[Lopez-Olivo et~al.(2015)Lopez-Olivo, Amezaga-Urruela, McGahan,
  Pollono, and Suarez-Almazor]{Lopez_Olivo_2015}
Maria~A. Lopez-Olivo, Matxalen Amezaga-Urruela, Lynda McGahan, Eduardo~N.
  Pollono, and Maria~E. Suarez-Almazor.
\newblock Rituximab for rheumatoid arthritis.
\newblock \emph{Cochrane Database of Systematic Reviews}, 1\penalty0
  (CD007356), 2015.
\newblock \doi{10.1002/14651858.CD007356.pub2}.

\bibitem[Rhodes et~al.(2020)Rhodes, Savovi\'{c}, Elbers, Jones, Higgins,
  Sterne, Welton, and Turner]{Rhodes_2020}
Kirsty~M. Rhodes, Jelena Savovi\'{c}, Roy Elbers, Hayley~E. Jones, Julian P.~T.
  Higgins, Jonathan A.~C. Sterne, Nicky~J. Welton, and Rebecca.~M. Turner.
\newblock Adjusting trial results for biases in meta-analysis: combining
  data-based evidence on bias with detailed trial assessment.
\newblock \emph{Journal of the Royal Statistical Society: Series A (Statistics
  in Society)}, 183\penalty0 (1):\penalty0 193--209, 2020.
\newblock \doi{10.1111/rssa.12485}.

\bibitem[Sahlin et~al.(2021)Sahlin, Troffaes, and Edsman]{Sahlin_2021}
Ullrika Sahlin, Matthias C.~M. Troffaes, and Lennart Edsman.
\newblock Robust decision analysis under severe uncertainty and ambiguous
  tradeoffs: an invasive species case study.
\newblock \emph{Risk Analysis}, May 2021.
\newblock \doi{10.1111/risa.13722}.

\bibitem[Spiegelhalter and Best(2003)]{Spiegelhalter_2003}
David~J Spiegelhalter and Nicola~G. Best.
\newblock Bayesian approaches to multiple sources of evidence and uncertainty
  in complex cost-effectiveness modelling.
\newblock \emph{Statistics in Medicine}, 22\penalty0 (23):\penalty0 3687--3709,
  2003.
\newblock \doi{10.1002/sim.1586}.

\bibitem[Stone et~al.(2020)Stone, Glass, Munn, Tugwell, and Doi]{Stone_2020}
Jennifer~C. Stone, Kathryn Glass, Zachary Munn, Peter Tugwell, and Suhail~A.R.
  Doi.
\newblock Comparison of bias adjustment methods in meta-analysis suggests that
  quality effects modeling may have less limitations than other approaches.
\newblock \emph{Journal of Clinical Epidemiology}, 117:\penalty0 36--45, 2020.
\newblock ISSN 0895-4356.
\newblock \doi{10.1016/j.jclinepi.2019.09.010}.

\bibitem[Troffaes(2017)]{Troffaes_2017}
Matthias C.~M. Troffaes.
\newblock A note on imprecise {M}onte {C}arlo over credal sets via importance
  sampling.
\newblock In Alessandro Antonucci, Giorgio Corani, In{\'e}s Couso, and
  S{\'e}bastien Destercke, editors, \emph{Proceedings of the Tenth
  International Symposium on Imprecise Probability: Theories and Applications},
  volume~62 of \emph{Proceedings of Machine Learning Research}, pages 325--332.
  PMLR, July 2017.
\newblock URL \url{http://proceedings.mlr.press/v62/troffaes17a.html}.

\bibitem[Troffaes(2018)]{Troffaes_2018}
Matthias C.~M. Troffaes.
\newblock Imprecise {M}onte {C}arlo simulation and iterative importance
  sampling for the estimation of lower previsions.
\newblock \emph{International Journal of Approximate Reasoning}, 101:\penalty0
  31--48, October 2018.
\newblock \doi{10.1016/j.ijar.2018.06.009}.

\bibitem[Turner et~al.(2009)Turner, Spiegelhalter, Smith, and
  Thompson]{Turner_2009}
Rebecca~M. Turner, David~J. Spiegelhalter, Gordon C.~S. Smith, and Simon~G
  Thompson.
\newblock Bias modelling in evidence synthesis.
\newblock \emph{Journal of the Royal Statistical Society. Serie A, (Statistics
  in Society)}, \penalty0 (1):\penalty0 21–47, 2009.
\newblock \doi{10.1111/j.1467-985X.2008.00547.x}.

\bibitem[{van der Bles} et~al.(2019){van der Bles}, {van der Linden}, Freeman,
  Mitchell, Galvao, Zaval, and Spiegelhalter]{vanderBles_2019}
Anne~Marthe {van der Bles}, Sander {van der Linden}, Alexandra~L.J. Freeman,
  James Mitchell, Ana~B. Galvao, Lisa Zaval, and David~J. Spiegelhalter.
\newblock Communicating uncertainty about facts, numbers and science.
\newblock \emph{The Royal Society}, 2019.
\newblock \doi{10.1098/rsos.181870}.

\bibitem[Verde(2021)]{Verde_2021}
Pablo~E. Verde.
\newblock A bias-corrected meta-analysis model for combining, studies of
  different types and quality.
\newblock \emph{Biometrical Journal}, 63\penalty0 (2):\penalty0 406--422, 2021.
\newblock \doi{10.1002/bimj.201900376}.

\bibitem[Walley(1997)]{Walley_1997}
Peter Walley.
\newblock Statistical inferences based on a second-order possibility
  distribution.
\newblock \emph{International Journal of General Systems}, 26\penalty0
  (4):\penalty0 337--383, 1997.
\newblock \doi{10.1080/03081079708945189}.

\bibitem[Welton et~al.(2009)Welton, Ades, Carlin, Altman, and
  Sterne]{Welton_2009}
N.~J. Welton, A.~E. Ades, J.~B. Carlin, D.~G. Altman, and J.~A.~C. Sterne.
\newblock Models for potentially biased evidence in meta-analysis using
  empirically based priors.
\newblock \emph{Journal of the Royal Statistical Society: Series A (Statistics
  in Society)}, 172\penalty0 (1):\penalty0 119--136, 2009.
\newblock \doi{10.1111/j.1467-985X.2008.00548.x}.

\bibitem[Xiong et~al.(2014)Xiong, Turner, Wei, Neal, Lyratzopoulos, and
  Higgins]{Tengbin_2014}
Tengbin Xiong, Rebecca~M. Turner, Yinghui Wei, David~E. Neal, Georgios
  Lyratzopoulos, and Julian P.~T. Higgins.
\newblock Comparative efficacy and safety of treatments for localised prostate
  cancer: an application of network meta-analysis.
\newblock \emph{BMJ Open}, 4\penalty0 (5), 2014.
\newblock \doi{10.1136/bmjopen-2013-004285}.

\end{thebibliography}

\appendix 

\section{Calculation of posterior expected overall effect}

The posterior expected overall effect is evaluated as:
\begin{equation}
\mathrm{E}_{\mathbf{q}}(\mu) = \idotsint \mu \cdot f_{\mathbf{q}}(\beta_1, \beta_2, \beta_3, \beta_4,\delta_1, \delta_2, \delta_3, \delta_4, \mu,\sigma^2_\theta|\mathbf{N}, \mathbf{r}, \mathbf{q})
\,d\beta_1 \,d\beta_2 \,d\beta_3 \,d\beta_4
\,d\delta_1 \,d\delta_2 \,d\delta_3 \,d\delta_4 \,d\mu \,d\sigma^2_{\theta} 
\end{equation}
where $\mathbf{N} = (N_{11}, N_{12}, N_{21}, N_{22}, N_{31}, N_{32}, N_{41}, N_{42})$ and $\mathbf{r} = (r_{11}, r_{12}, r_{21}, r_{22}, r_{31}, r_{32}, r_{41}, r_{42})$
are the total number of patients and the number of patients who have had a positive response in control and treatment groups in each study respectively (i.e. data),
$\mathbf{q} = (q_1, q_2, q_3, q_4)$ are study qualities and
\begin{multline}
f_{\mathbf{q}}(\beta_1, \beta_2, \beta_3, \beta_4,\delta_1, \delta_2, \delta_3, \delta_4, \mu,\sigma^2_\theta|\mathbf{N}, \mathbf{r}, \mathbf{q}) \\ \propto \prod_{i=1}^4\prod_{j=1}^2 g^{-1}(\eta_{ij})^{r_{ij}} \Big(1 - g^{-1}(\eta_{ij})\Big)^{N_{ij} - r_{ij}} \cdot  f(\beta_i) \cdot  f_{q_i}(\delta_i|\mu,\sigma^2_\theta) \cdot  f(\mu) \cdot  f(\sigma^2_\theta)
\end{multline}
is the posterior distribution for given study quality ${\mathbf{q}}$, $g(x) = \mathrm{logit}(x)$,
$\eta_{i1} = \beta_i$,  $\eta_{i2} = \beta_i + \delta_{i}$, and $f(\beta_i)$, $f_{q_i}(\delta_i|\mu,\sigma^2_\theta)$, $f(\mu)$ and $f(\sigma^2_\theta)$ are previously specified prior distributions for parameters $\beta_i$, $\delta_i$, $\mu$ and $\sigma^2_\theta$ respectively.

\section{Additional forestplots}

\begin{figure}
  \begin{subfigure}{0.95\textwidth}
      \centering
      \includegraphics[width=\textwidth]{Forestplot_D1.pdf}\label{fig:Forestplot_D1_again}
 \end{subfigure}
  \begin{subfigure}{0.95\textwidth}
      \includegraphics[width=\textwidth]{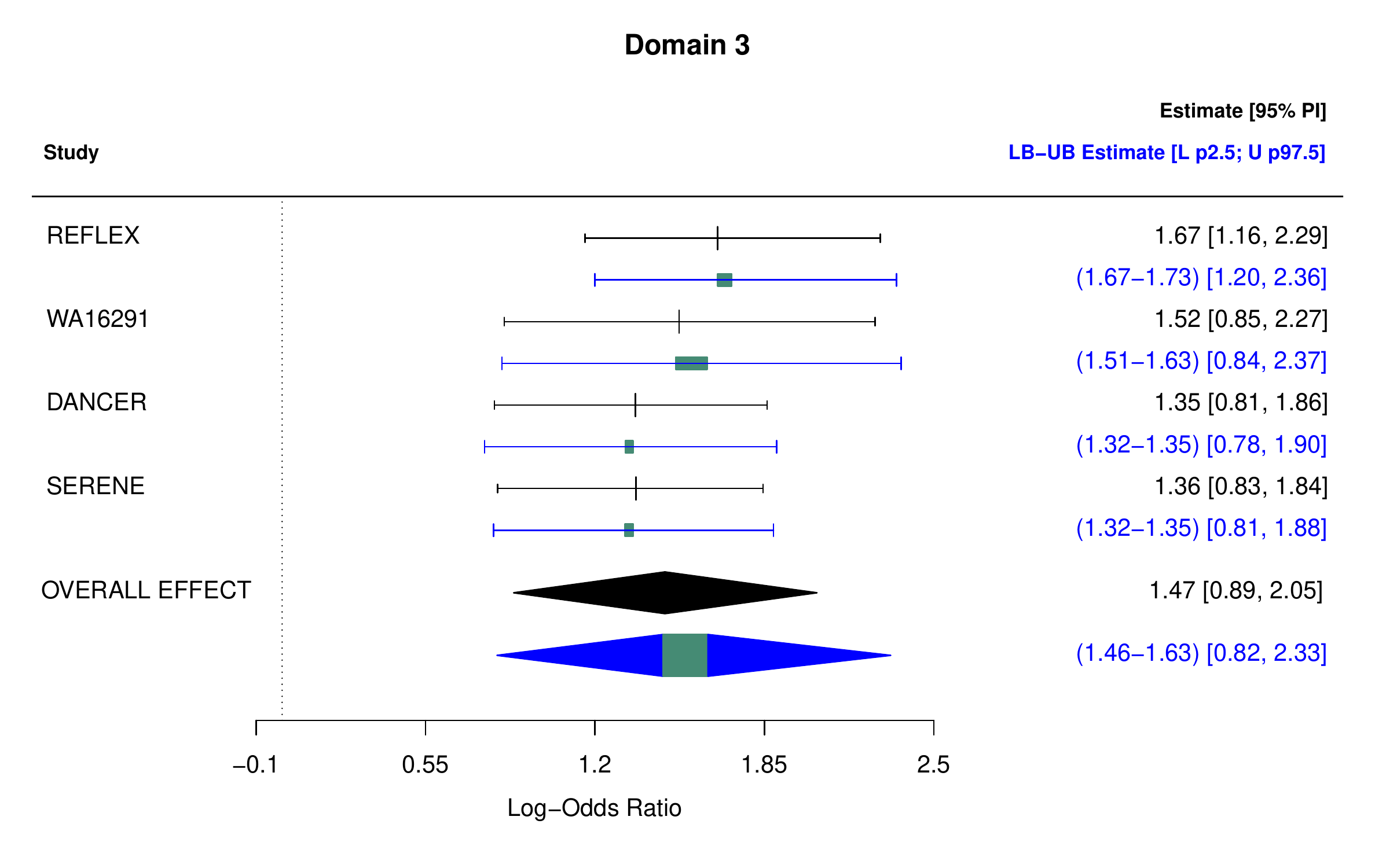}\label{fig:Forestplot_D3}
  \end{subfigure}
   \caption{Forestplot of a meta-analysis of the effectiveness of Rituximab plus metrotexato modified to show bounds on quantities of interest. Unadjusted and robust Bayesian bias-adjusted random effects log-odds ratios (with 95\% PI) are displayed: (black) unadjusted model; (blue) robust bias-adjusted random effects model. For the robust bias-adjusted random effects model, bounds on the expected overall effect, the lower 2.5th percentile and the upper 97.5th percentile are shown.}\label{fig:Forestplot}
\end{figure}

\begin{figure}
  \begin{subfigure}{0.95\textwidth}
      \includegraphics[width=\textwidth]{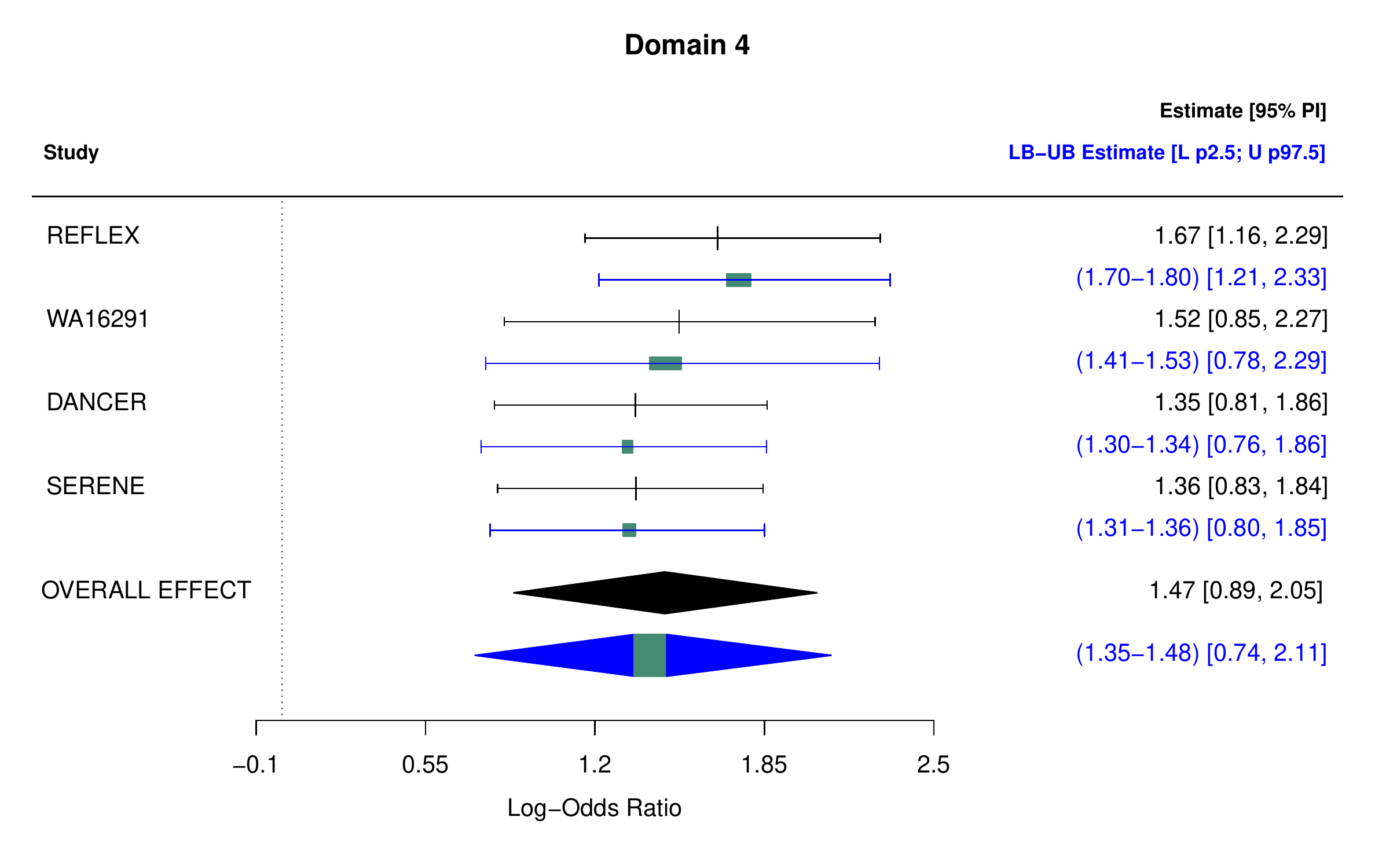}\label{fig:Forestplot_D4}
  \end{subfigure}
  \begin{subfigure}{0.95\textwidth}
      \includegraphics[width=\textwidth]{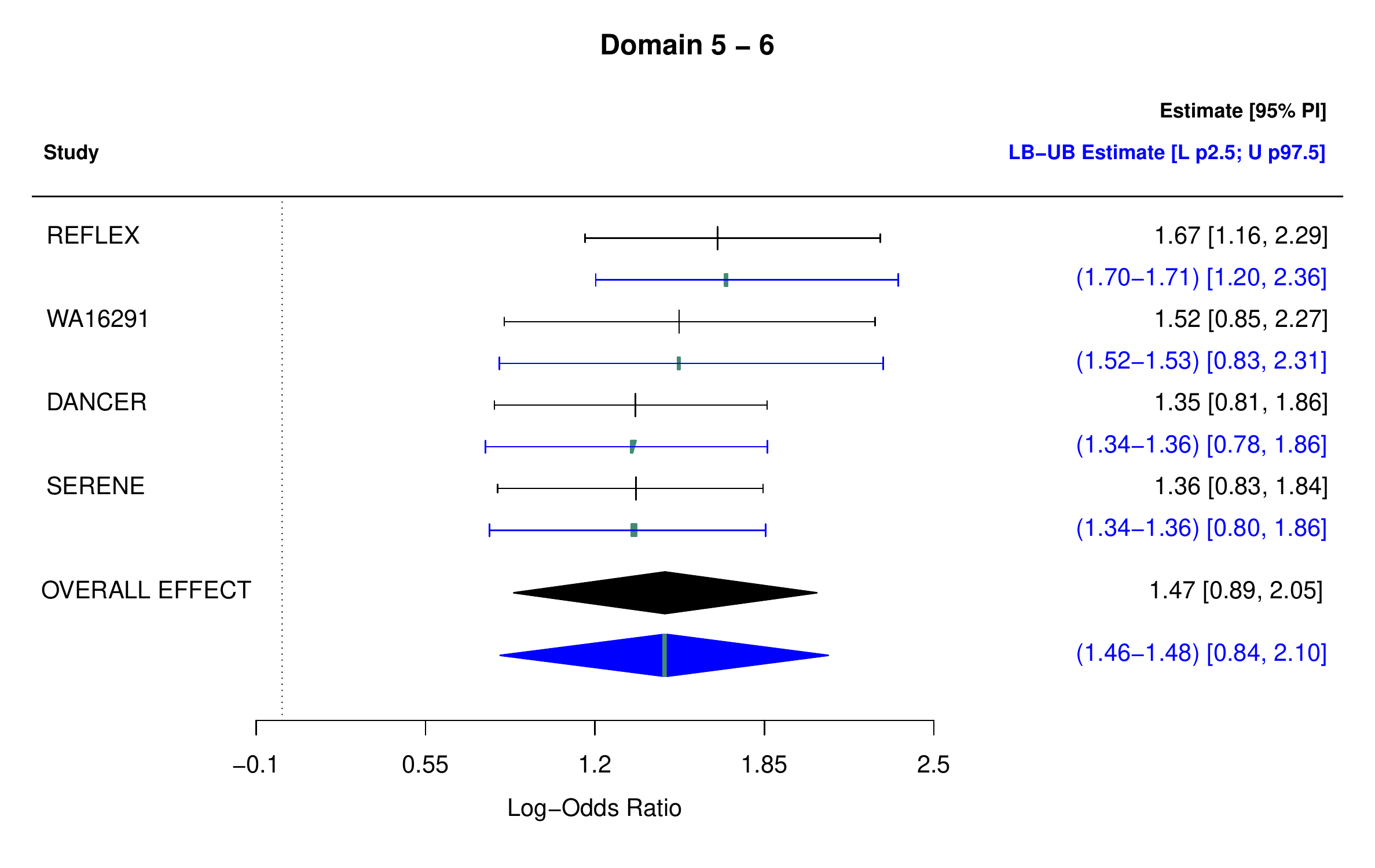}\label{fig:Forestplot_D5}
  \end{subfigure}
     \caption{Forestplot of a meta-analysis of the effectiveness of Rituximab plus metrotexato modified to show bounds on quantities of interest. Unadjusted and robust Bayesian bias-adjusted random effects log-odds ratios (with 95\% PI) are displayed: (black) unadjusted model; (blue) robust bias-adjusted random effects model. For the robust bias-adjusted random effects model, bounds on the expected overall effect, the lower 2.5th percentile and the upper 97.5th percentile are shown.}\label{fig:Forestplot_D_4_5_6}
\end{figure}

\begin{figure}
  \begin{subfigure}{0.95\textwidth}
      \includegraphics[width=\textwidth]{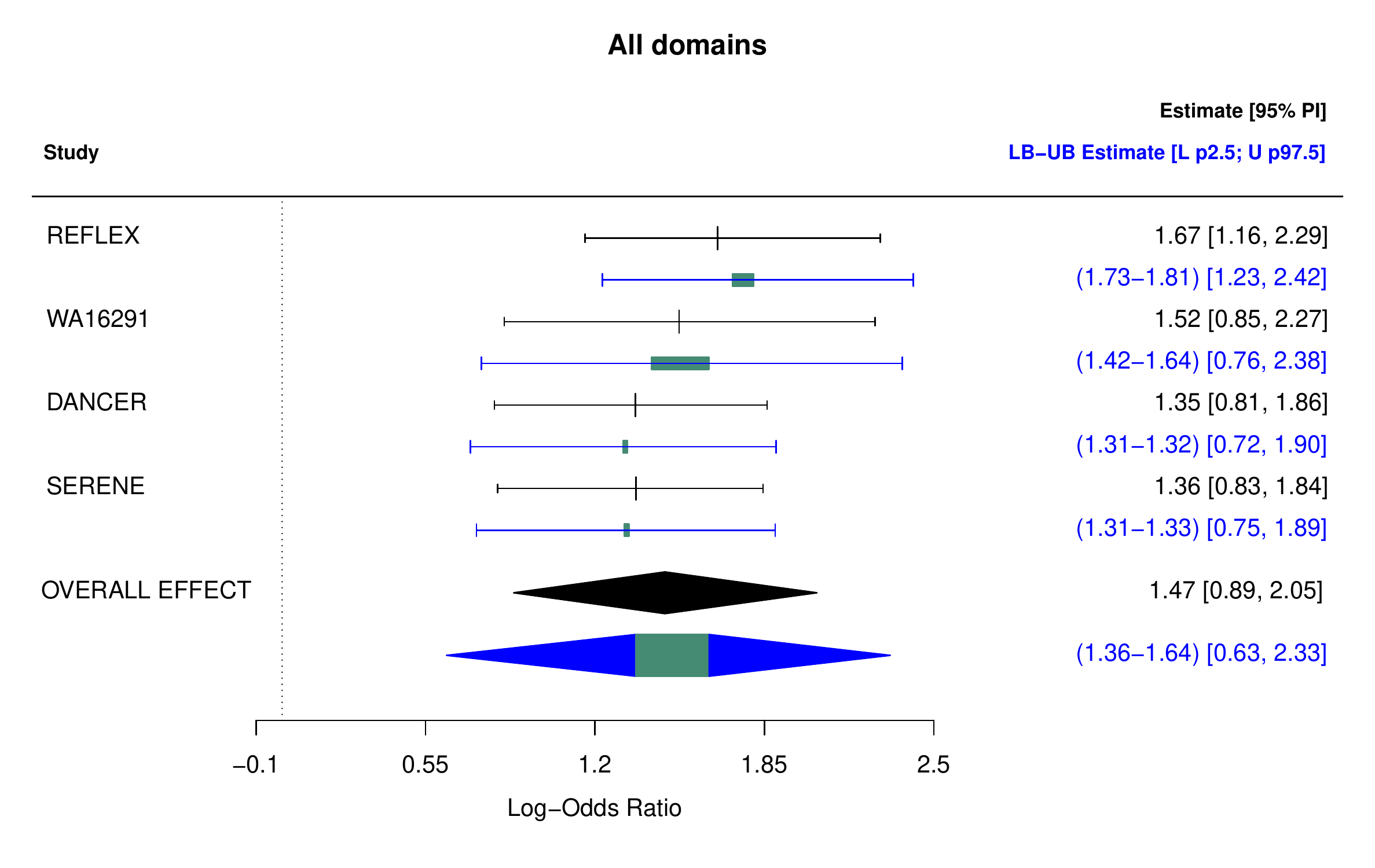}\label{fig:Forestplot_all_domains}
  \end{subfigure}
     \caption{Forestplot of a meta-analysis of the effectiveness of Rituximab plus metrotexato modified to show bounds on quantities of interest. Unadjusted and robust Bayesian bias-adjusted random effects log-odds ratios (with 95\% PI) are displayed: (black) unadjusted model; (blue) robust bias-adjusted random effects model. For the robust bias-adjusted random effects model, bounds on the expected overall effect, the lower 2.5th percentile and the upper 97.5th percentile are shown.}\label{fig:Forestplot_all}
\end{figure}

\Cref{fig:Forestplot,fig:Forestplot_D_4_5_6,fig:Forestplot_all} show the results of the random effects model with and without robust Bayesian bias-adjustment, for each risk of bias domain.

\end{document}